\documentclass[aps, prd, twocolumn, lengthcheck, superscriptaddress, showpacs, letterpaper, nofootinbib]{revtex4-1}

\usepackage{epsfig}
\usepackage[usenames]{color}
\usepackage{graphicx}
\usepackage{amsmath}
\usepackage{epstopdf}
\usepackage{multirow}
\usepackage{ulem}
\newcommand{\sbar}[1]{\ooalign{\hfil/\hfil\crcr$#1$}}

\def\bi{\bibitem}

\def\la{\langle}\def\ra{\rangle}
\def\be{\begin{eqnarray}}\def\ee{\end{eqnarray}}
\def\lsim{\mathrel{\rlap{\lower3pt\hbox{\hskip1pt$\sim$}}
     \raise1pt\hbox{$<$}}} 
\def\gsim{\mathrel{\rlap{\lower3pt\hbox{\hskip1pt$\sim$}}
     \raise1pt\hbox{$>$}}} 
\def\del{\partial}

\allowdisplaybreaks

\def\la{\langle}\def\ra{\rangle}
\def\be{\begin{eqnarray}}\def\ee{\end{eqnarray}}
\def\lsim{\mathrel{\rlap{\lower3pt\hbox{\hskip1pt$\sim$}}
     \raise1pt\hbox{$<$}}} 
\def\gsim{\mathrel{\rlap{\lower3pt\hbox{\hskip1pt$\sim$}}
     \raise1pt\hbox{$>$}}} 
\def\le{ \begin{array}{ll}}\def\re{\end{array}}

\def\lear{ \left( \begin{array}{cc}}\def\rear{\end{array} \right)}
\def\tr{\textnormal{tr}}

\def\le{ \left( \begin{array}{cc}}\def\re{\end{array} \right)}
\def\tr{\textnormal{tr}}

\def\bi{\bibitem}
\def\bchi{\bar{\chi}}

\def\eft-hls{{\it EFT}$_{\rm bsHLS}$}
\def\skyrmion-hls{{\it Skyrmion}$_{\rm sHLS}$}

\def\Sigmabar{$\overline{\Sigma}$}
\def\Lamb-chiral{\Lambda_{\rm chiral}}
\def\Lamb-bshls{\Lambda-{\rm bshls}}
\def\Z{{\cal Z}}

\begin{document}

\title{Pseudoconformal structure in dense baryonic matter}
\author{Yong-Liang Ma}
\email{yongliangma@jlu.edu.cn}
\affiliation{Center for Theoretical Physics and College of Physics, Jilin University, Changchun, 130012, China}

\author{Mannque Rho}
\email{mannque.rho@ipht.fr}
\affiliation{Institut de Physique Th\'eorique, CEA Saclay, 91191 Gif-sur-Yvette c\'edex, France }

\date{\today}

\begin{abstract}
We clarify and extend further the idea we developed before that baryonic matter at high density has an emergent ``pseudoconformal symmetry."  It is argued that as baryonic density exceeds {$n \simeq n_{1/2} \gtrsim 2n_0$}, a topology change mimicking the baryon-quark continuity takes place at $n_{1/2}$. In terms of skyrmions, this  corresponds to the transition from skyrmions to half-skyrmions and impacts on the equation of state of dense baryonic matter. The emergence in medium at $n_{1/2}$ of parity-doublet symmetry --- which is invisible in QCD in a matter-free vacuum --- plays the crucial role. The consequence {of the topology change} is that massive compact stars carry the ``pseudoconformal sound velocity" $v_s^2/c^2\approx 1/3$ at {$n\gsim n_{1/2}$} signaling a precursor to the precocious emergence of scale symmetry as well as {a} local symmetry hidden in QCD in the matter-free vacuum. A highly significant prediction of this work is that the topology change density from normal matter to half-skyrmion matter, up to date inaccessible either by QCD proper or by terrestrial experiments, could possibly be pinned down within the range $2 < n_{1/2}/n_0 < 4$, commensurate with the range expected for a continuous hadrons-to-quarks or -gluons transition.
\end{abstract}

\maketitle

\section{ Introduction}

\subsection{Objective}

Baryonic matter above the nuclear equilibrium density $n_0\sim 0.16$ fm$^{-3}$, say, $n\gsim 2n_0$, is totally unknown both theoretically and experimentally and has been a long-standing challenge to nuclear physicists. The most glaring case for this situation is aptly illustrated in the nuclear symmetry energy that enters in the equation of state (EOS) of asymmetric baryonic matter relevant to (neutron-rich) compact stars.  While constrained by experiments up to $\sim n_0$, the symmetry energy given by available theoretical models can wildly vary above $n_0$, presenting a total wilderness, and the available experiments, mainly coming from heavy-ion experiments, are unable to weed out the wilderness. The lattice approach, presently the only nonperturbative tool in QCD, cannot access the density regime involved.

The recent LIGO/Virgo detection of gravitational waves emitted from coalescing neutron stars GW170817~\cite{GW170817} opens a new era of nuclear physics. With such ongoing and upcoming detections, it is reasonable to expect that astronomical observations will bring answers to outstanding problems in both astrophysics and nuclear physics, such as, for example, the structure of highly dense matter in compact stars and ultimately the origin of proton mass.

In this paper, we follow the strategy first proposed in \cite{PKLMR,MLPR}  to access high baryon density in a single field theoretic framework involving {\it only hadronic degrees of freedom}. It exploits topology, scale symmetry and local flavor symmetry, all invisible in QCD in the vacuum. The underlying idea developed is basically different from what is currently adopted in the nuclear-astrophysics community.

Topology enters at high density in the framework in place of quarks and gluons,  dilatonic scalar figures signaling the precursor to the emergence of scale symmetry broken in the vacuum  and the vector meson $\rho$, embodying hidden flavor local symmetry, manifests explicitly at what is called the vector manifestation (VM) fixed point. At the vector manifestation fixed point, the vector meson mass goes to zero, exposing local flavor symmetry which is absent in QCD in the vacuum. At a certain high density, a topology change takes place  and affects profoundly the properties of neutron stars, such as the sound velocity and the tidal deformability~\cite{MLPR}. The purpose of the present work is to sharpen and extend the arguments developed in the previous work~\cite{PKLMR,MLPR}.

\subsection{Theoretical tool}

Our principal tool is the  effective Lagrangian --- and, more broadly, the approach anchored on  it --- called $bs$HLS that incorporates, in addition to the (pseudo-)Goldstone bosons, $\pi$, and the nucleons that figure in standard chiral perturbation theory (S$\chi$PT), the lowest-lying vector mesons $\rho$ and $\omega$ and a scalar meson, dilaton,  $\chi$ (or $\sigma$).  As formulated in detail in \cite{PKLMR} (and references cited therein) and briefly reviewed below, the parameters of the effective field theory (EFT) Lagrangian are endowed with various QCD condensates such as $\langle \bar{q}q \rangle, \langle G^2 \rangle$, etc., brought in by the matching between the EFT and QCD  via correlators~\cite{HY:WilsonMatch}. Given that the condensates depend on the vacuum involved, and in a nuclear medium, the vacuum is changed by density, the bare parameters of the Lagrangian must thereby depend on density. This dependence is referred to as ``intrinsic density dependence (IDD)" as opposed to density dependence induced by nuclear correlations, called ``induced density dependence," that needs to be taken into account in doing renormalization decimations.

\subsection{Outcome}

The consequences are  striking. Even though the trace of the energy-momentum tensor (TEMT) is found not to be zero, the sound velocity in massive stars is found to converge quite precociously to $v_s^2/c^2\approx 1/3$, what is commonly associated with conformal invariance. Combined with currently available information from heavy-ion collisions and from stellar observations such as the tidal deformability inferred from gravitational waves and accurately measured maximum star masses, the topology change density, accessible neither experimentally nor theoretically,  can be pinned down to  within the range
 \be
 2 < n_{1/2}/n_0 < 4.\label{bound}
 \ee

This paper is organized as follows: In Sec.~\ref{sec:topology}, we discuss the role of topology  in chiral effective theory and the impact of topology change on nuclear dynamics in dense matter. In Sec.~\ref{sec:hidden}, we specify the hidden symmetries of QCD involved and their role in  the effective field theory $bs$HLS. We devote ourselves in Sec.~\ref{sec:DLFP} to discuss how to access in $bs$HLS the dilaton-limit fixed point at which the symmetries, i.e., parity doubling, scale symmetry and flavor local symmetry, are revealed. We then obtain the equation of state of the pseudoconformal nuclear matter in Sec.~\ref{sec:PCM} and study the compact-star properties in Sec.~\ref{sec:star}.  The outcome of the analysis is that the density at which the topology change takes place could be pinned down.  Our further discussions and perspectives are given in Sec.~\ref{sec:dis}.

\section{Topology}

\label{sec:topology}

\subsection{Return of ``Cheshire cat"}

Although not rigorously proven, it is generally considered most likely  that at some high density going above $n_0$, there should be a change of degrees of freedom from hadrons to those of QCD, namely, quarks and gluons.  The changeover seems to be indispensable in compact stars for accounting for the $\gsim 2$-solar-mass stars and more specifically in the approach we are presenting in this paper for what we call ``pseudoconformal" sound velocity of stars $v^2_s/c^2\simeq 1/3$  that sets in precociously at $n\gsim 3n_0$. In this paper we approach this changeover in terms of topology change as  a ``trade-in"  from baryons to strongly coupled quarks. The idea is motivated by the Cheshire cat mechanism~\cite{cheshirecat} given by the chiral bag model for the baryons~\cite{chiralbag}.

The basic premise of the chiral bag model (CBM) is anchored on the observation that the spontaneous breaking of chiral symmetry in QCD is tied to quark confinement~\cite{casher}.  Although confinement is not very well understood, the CBM relates in terms of a bag model the manifestation of broken chiral symmetry with quarks and gluons confined inside  a bag --- called the MIT bag --- and Goldstone bosons ($\pi$) (and heavier mesons) living outside of the bag coupled to the quarks and gluons inside by suitable boundary conditions. What is crucial in this model is that the Goldstone boson (pion) can carry a baryon charge in the guise of a  skyrmion~\cite{Skyrme} and renders the bag  as a ``gauge artifact."  It has transpired in the decades of developments that physics at low energy should be independent of the confinement size, say, bag radius $R$. Thus, for example, a baryon can be equally described as the MIT bag ($R=\infty$) as well as the skyrmion ($R=0$). This is the Cheshire cat (CC) principle~\cite{cheshirecat}.  Lacking an exact bosonization technique in (3+1) dimensions, this is only an approximation --- perhaps too drastic --- in nature except for the topological charge, i.e., baryon charge. It is technically involved to do fully rigorous calculations to check how the notion of the Cheshire cat phenomenon works in hadron dynamics. It has been, however, looked at in detail for the flavor-singlet axial charge (FSAC) of the proton $g_A^0$. It has been shown that  the FSAC is indeed independent of the confinement size~\cite{fsac}. This quantity involves an intricate interplay  of $U_A(1)$ anomaly with color boundary conditions, so a highly nontrivial manifestation of the CC phenomenon.

\subsection{Topology change}

We now propose to apply an argument along the line of the Cheshire cat mechanism to implement the possible change of degrees of freedom as the density goes above $n_0$ to the range of densities relevant to compact stars, viz. $\sim (5-7)n_0$. Rough consideration based on the increase of hadronic size at increasing density and the onset of quark percolation suggests that the quark or gluon degrees of freedom could enter at $\sim 2n_0$. For example, a very likely scenario consistent with the existence of $\sim 2$-solar-mass stars posits the intervention of strong-coupled quark matter in a smooth hadron-quark transition at $\sim (2{\rm-}4)n_0$ with the perturbative QCD effect setting in at $\gsim 10 n_0$ and, ultimately, the color-flavor locking (CFL) at $\sim100 n_0$~\cite{baymetal}.\footnote{There have been numerous works in which the quark degrees of freedom are implemented {\it explicitly}, typically with phase transitions, in the equation of state for compact stars, among which notable is Ref.~\cite{Quarks}. It will be seen in what follows that our approach differs basically from what is involved in those approaches.  In particular, the transition involved will not be of the Landau-Ginzburg-Wilsonian paradigm of phase transitions. We will elaborate on this point in Sec.~\ref{last}.}
In this paper, we propose that the intervention of the quark or gluon degrees of freedom be captured by a topology change as a possible implementation of the continuous hadron-quark continuity.  For our purpose, it seems more appropriate to consider the quark-hadron continuity in the form of quarkyonic matter applied to compact stars~\cite{kojo,mclerran}.

The difference from the quark-hadron continuity is that topology change involves ``hadronic" variables only whereas in the latter {\it explicitly} two different variables figure. It is worth pointing out here that the quarkyonic matter could be ``baryonic"~\cite{philipsen}  as will be discussed further later.

We will be concerned with densities of order $\lsim 10 n_0$, so we will not venture into the CFL regime. However as shown in \cite{qualiton}, it is feasible to phrase even the CFL phase involving asymptotic densities in terms of hadronic variables, e.g., ``pions," ``vector mesons," etc.  We think it should be possible to formulate this picture in terms of a Cheshire cat phenomenon.

The topology change we will consider is natural in QCD at large $N_c$ and large density. At that limit, the baryonic matter, QCD suggests,  must be in the form of crystal populated most likely by half-skyrmions. There is a hint for this in the Sakai-Sugimoto holographic QCD model~\cite{SS}, which is thought to represent QCD in that limit~\cite{dyonicsalt}. Numerical simulation  shows indeed that skyrmions put on a face-centered-cubic (fcc) crystal lattice change over to half-skyrmions in CC as the lattice size is reduced, corresponding to the increase of density. The topology change density is referred to as $n_{1/2}$.  For a review see \cite{park-vento}.  The numerical value for $n_{1/2}$ cannot be pinned down from the skyrmion model because it depends on various quantities, such as the degrees of freedom included in the Lagrangian, but the transition itself, being topological, is considered to be robust. From here on we will assume this robustness in our discussions.

What takes place in the changeover here, loosely called transition throughout this paper,  is not a bona fide Landau-Ginzburg-type phase transition since it involves no local order parameter. At the transition we are concerned with, what corresponds to $\Sigma\equiv \la\bar{q}q\ra$ --- called quark condensate ---  goes to zero when space-averaged, $\overline{\Sigma}\to 0$, but is not zero locally, bearing inhomogeneity~\cite{inhomogeneous}. Thus the vanishing of  $\overline{\Sigma}$ does not signal chiral symmetry restoration. There are pions present, so the symmetry is still broken. This implies that $\Sigma$ is not an order parameter of chiral symmetry. There must then be something else that represents the order parameter of chiral symmetry. It is presently unknown what that is~\cite{stern}.\footnote{That it could be a four-quark condensate has been discussed. There are arguments however that such four-quark condensates as an order parameter are ruled out by a refined 't Hooft anomaly matching even in the presence of density~\cite{tanizaki}.} The situation is somewhat like a pseudogap in superconductivity~\cite{pseudogap}.\footnote{The possible analogy to superconductivity~\cite{pseudogap}: $\overline{\Sigma}$ is the analog to the spectral gap $\Delta_s$ and the pion decay constant $f_\pi$ to the (pairing) order parameter $\Delta_0$.}

As discussed in \cite{cho}, the half-skyrmions that appear at $n\geq n_{1/2}$ are confined to a skyrmion, so they are not propagating objects.  The key observation is that there are hidden gauge symmetries in the chiral field $U=e^{2i\pi/f_\pi}$ that figures in the skyrmion Lagrangian. Apart from the hidden (non-Abelian) local symmetry discussed below, it has a hidden local $U(1)$ symmetry. To see this, let us consider the hedgehog ansatz for the static chiral field, $U_0(r)$,
\be
U_0 (r)=e^{i\vec{\tau}\cdot\hat{r} \theta(r)}
\ee
where $\theta (r)$ is the chiral angle that goes from 0 to $\pi$. When $\hat{r}$ is parameterized by the CP$^1$ field $z$ with $z^\dagger z=1$,
\be
\hat{r}=z^\dagger \sigma z,
\ee
then $U_0$ is invariant under the $U(1)$ gauge transformation $z\to e^{i\kappa (r)} z$. Elevating this redundancy to the $U(1)$ gauge field, one finds the skyrmion theory as the $U(1)$ gauge field coupled to a massless scalar field. There is then a ``hidden" monopole in the theory and the regularized monopole can be identified with the skyrmion~\cite{cho,Nitta}.

Furthermore there are other monopole solutions. Among them there are monopole and half-skyrmion solutions that have  infinite energy at infinity when separated but bound to finite-energy skyrmions. Thus one can think of a skyrmion as a monopole of confined half-skyrmions.\footnote{This configuration is different from what is thought to  happen in the N\'eel-VBS (valence bond solid) transition in (2+1) dimensions where the monopole event is suppressed by a Berry phase and hence the half-skyrmions are deconfined in what is known as a ``deconfined quantum critical phenomenon"~\cite{senthil}.}

Given that they are confined, the  object of half-skyrmions is a {\it  modified} baryon. The half-skyrmions could very well be present in nuclei. Indeed even an $\alpha$ particle can be reasonably considered as a complex of eight half-skyrmions~\cite{Manton:2017bee}. {\it What distinguishes the state of matter for density $n<n_{1/2}$ from that of $n\geq n_{1/2}$ is that $\overline{\Sigma}$, nonzero in $n<n_{1/2}$, goes to zero at $n=n_{1/2}$ while $f_\pi$ stays nonzero across $n_{1/2}$.}

\subsection{Nuclear symmetry energy}

A striking effect of the topology change at high density is the dramatic change in the structure of the nuclear symmetry energy. This was first seen in dense skyrmion matter simulated on a crystal lattice~\cite{cusp}. The symmetry energy for baryonic matter, $E_{sym}$,  that figures in the energy per nucleon of the baryonic matter $E(n)$ at density $n$ as
\be
E(n, \alpha)= E(n,\alpha=0) +E_{sym} (n) \alpha^2 + O(\alpha^4),
\ee
where $\alpha=(N-Z)/(N+Z)$ with $N(Z)$ being the number of neutrons (protons) in the system, plays the key role in the EOS of compact-star matter. The $E_{sym}$ can be obtained from the skyrmion matter by collective-quantizing the pure neutron matter (i.e., $\alpha=1$). It is given by $E_{sym}=1/(8{\cal I}) + O(1/N_c)^2$, where ${\cal I}$ is the isospin moment of inertia~\cite{cusp}.  At high density and large $N_c$ we expect this result to be justified. Being topological it should be robust. A surprising result found in \cite{cusp} is that since $\overline{\Sigma}$ goes to zero as density approaches $n_{1/2}$, there appears a cusp at $n_{1/2}$, with the symmetry energy dropping going toward $n_{1/2}$, then turning over and increasing as the density increases beyond $n_{1/2}$. This cusp is highly robust against strong interactions~\cite{robust} but, being semiclassical, is expected to be smoothed by higher order nuclear correlations. Also the pion mass will intervene in eliminating discontinuity in the cusp structure. In \cite{PKLMR}, this cusp was reproduced in terms of the tensor force constructed with the Lagrangian. This cusp structure can be simply understood in terms of the behavior of the nuclear tensor force~\cite{cusp}.

\subsection{Parity-doublet structure in topology change}

The topology change that drives \Sigmabar,\  nonzero in the skyrmion phase simulated in fcc,  to zero in CC at $n_{1/2}$ exposes parity-doublet structure in baryonic matter~\cite{maetal-doublet}. We will see below that this will play a crucial role for the pseudoconformal structure.

On a crystal lattice, the effective pion decay constant $f_\pi^\ast$ encoding the IDD (precisely defined below in Sec.~\ref{bshls}) is found
to drop smoothly as the lattice is reduced, roughly in consistency with chiral
perturbation theory, toward the density $n_{1/2}$, but stops
dropping at $n_{1/2}$ where half-skyrmions appear and remains constant $ \sim (60-80)\%$ of the free-space value in the half-skyrmion phase~\cite{Lee:2003aq}.
The in-medium nucleon mass $m_N^\ast$ tracks closely the
in-medium pion decay constant $f_\pi^\ast$ multiplied by a
scale-invariant factor proportional to $\sqrt{N_c}$ which
indicates that the large $N_c$ dominance holds in the
medium as it does in free space and stays constant~\cite{maetal-doublet}. Given that $\overline{\Sigma}\to 0$ in the half-skyrmion phase, we associate this constant
 with the chiral-invariant mass $m_0$ that we will encounter in
the parity-doublet baryon model discussed below in Sec.~\ref{sec:DLFP}. It is significant that such a chirally invariant term is not explicitly present in
the Lagrangian with which the skyrmion crystal is constructed and hence must be generated by nuclear correlations.

\section{Hidden symmetries of QCD}

\label{sec:hidden}

In order to proceed to implement the topology change described above in an EFT framework, we exploit symmetries of QCD that are not visible in the matter-free vacuum that could emerge in a dense medium through strong nuclear correlations. We focus on two, one, scale symmetry and the other, hidden local symmetry {(HLS)}. Since this matter is discussed in detail elsewhere~\cite{PKLMR,MLPR}, we briefly summarize only the key points relevant to our line of arguments developed.

\subsection{Flavor local symmetry}

{ QCD has no flavor (local) gauge symmetry but the chiral field $U=e^{2i\pi/f_\pi}$ in the effective $SU(N_f)_L\times SU(N_f)_R$ chiral Lagrangian has redundancies. For instance, written in terms of $L$(eft) and $R$(ight) fields,  one such (local) redundancy is $h^\dagger(x) h((x)=1$ inserted between the chiral $R$ and $L$ fields as
\be
U(x)= \xi_L^\dagger (x)\xi_R (x) =\xi_L^\dagger (x) h(x)^\dagger h(x) \xi_R (x).
\ee
There can of course be infinite such redundancies sandwiched between the $L$ and $R$ fields.
These redundancies can be elevated to gauge symmetries by introducing gauge fields. If kinetic energy terms are generated by dynamics, then they can give rise to a local gauge theory consisting of an infinite tower of vector mesons. How this can actually happen in hadronic dynamics was recently discussed for one redundancy in \cite{yamawaki2018}.\footnote{ It is perhaps not recognized in the nuclear theory community working on chiral effective field theory with pions only that local gauge symmetry is in fact in the theory but hidden in higher chiral-order terms. That one can write down a hidden gauge symmetric Lagrangian with the vector manifestation fixed point {\it \`a la} Yamawaki~\cite{yamawaki2018} implies that the vector mesons can be thought of as emergent-symmetry fields from a nonlinear sigma model (say, in the chiral limit)~\cite{suzuki,barcelo}. This implies that looking in relativistic heavy-ion experiments for what was thought to be ``Brown-Rho" scaling  was a ``wild goose chase" in the wrong places. }

In string theory, such an infinite tower of hidden local gauge fields does arise from 5D Yang-Mills theory~\cite{SS} and can account for the vector dominance structure of the nucleon EM form factors~\cite{VD}. For our problem restricted to two flavors ($N_f=2$) defined at the chiral scale, we do not need all the high tower of vector fields, so we focus on the lowest vector mesons $\rho$ and $\omega$ with $\rho\in SU(2)$ and $\omega\in U(1)$, integrating out the higher members of the tower. } As explained below, the flavor $U(2)$ symmetry for the vector mesons, fairly good at low density,  is strongly violated at high density, so they will be treated separately. Thus we will be primarily interested in
\be
h(x)\in SU(2)_{L+R}\times U(1)_{L+R}.
\ee

Written in terms of one-forms
\begin{eqnarray}
\hat{\alpha}_{\parallel\mu} & = & \frac{1}{2i}(D_\mu \xi_R \cdot
\xi_R^\dag +
D_\mu \xi_L \cdot \xi_L^\dag), \nonumber\\
\hat{\alpha}_{\perp\mu} & = & \frac{1}{2i}(D_\mu \xi_R \cdot
\xi_R^\dag - D_\mu \xi_L \cdot \xi_L^\dag)
\ , \label{eq:1form}
\end{eqnarray}
where the covariant derivative is defined as $D_\mu \xi_{R,L} = (\partial_\mu - i V_\mu)\xi_{R,L}$ with
\be
V_\mu(x) & = & \frac{g_{\rho}}{2}\rho_\mu^a \tau^a + \frac{g_{\omega}}{2}\omega_\mu I_{2\times 2},
\ee
the HLS Lagrangian to the leading order in the power counting is
\be
{\mathcal L}_M
& = & {f_{\pi}^2}\mbox{tr}\left[ \hat{\alpha}_{\perp\mu}
  \hat{\alpha}_{\perp}^{\mu} \right]
{}+ a_\rho f_{\pi}^2\mbox{tr}\left[ \hat{\alpha}_{\parallel\mu}
  \hat{\alpha}_{\parallel}^{\mu} \right]\nonumber\\
&&{}+ (a_\omega-a_\rho) f_{\pi}^2 \mbox{tr}\left[ \hat{\alpha}_{\parallel\mu} \right]
  \mbox{tr}\left[ \hat{\alpha}_{\parallel}^{\mu} \right] \nonumber\\
 &&{}- \frac{1}{2}\mbox{tr}\left[ \rho_{\mu\nu}\rho^{\mu\nu} \right]
{}- \frac{1}{2}\mbox{tr}\left[ \omega_{\mu\nu}\omega^{\mu\nu} \right]\,.
\ee
Here $a_V$ for $V=(\rho,\omega)$ is a parameter that enters in the mass formula $m_V^2=a_V f_\pi^2 g_V^2$~\cite{HY:PR}.  For the $\rho$ meson, $a_\rho=2$ gives the familiar Kawarabayashi-Suzuki-Riazuddin-Fayyazuddin (KSRF) formula.
Although this Lagrangian is fairly well known, we have written it down explicitly here for definition of the terms to be used below.

Let us ignore for the moment the $\omega$ meson and consider only the $\rho$. We will come to the $\omega$ matter later since its property is important for the role of parity-doublet symmetry.

The advantage of treating the $\rho$ as a hidden local symmetric field is that it can be considered on the {\it same footing} as the pion, that is, the $\rho$ mass being in some sense as ``light" as the pion mass. In fact treating it as such, one can do a systematic chiral perturbation calculation as shown in \cite{HY:PR}. This makes a good  sense in dense matter as one expects the $\rho$ mass to go down at increasing density. The  KSRF mass formula, successful in the matter-free vacuum,  becomes a lot more accurate as the hidden gauge coupling $g_\rho$ falls along with the chiral condensate. In fact, Wilsonian RG analysis shows that the gauge coupling should go to zero --- hence  also the mass --- as the density approaches what is called the VM fixed point~\cite{HY:VM1,HY:VM2}. Note that the mass goes to zero not because the order parameter of chiral symmetry $f_\pi$ goes to zero at high density but because the $\rho$-nucleon coupling goes to zero.\footnote{ It will be seen below (in Sec. \ref{sec:DLFP}) that the $\rho$ meson could decouple from nucleons {\it before} the VM fixed point is arrived at. This means that the $\rho$ mass in dense medium can drop to zero even if the gauge coupling $g_\rho\neq 0$ and $f_\pi\neq 0$.  As pointed out in a footnote above, this may have a different chiral symmetry property in a dense medium from high-temperature systems where the $\rho$ mass goes to zero due to the VM fixed point.} It is at the point where the $\rho$ decouples approaching the VM fixed point that the local gauge symmetry for the $\rho$ manifests. This symmetry is not in QCD in the matter-free vacuum. Hence it is appropriate to view it as a symmetry ``emerging" in the system as the quark condensate is driven to zero. The VM fixed point is estimated to lie at $n\gsim 20n_0$~\cite{PKLMR}.

If the $\omega$ were considered in $U(2)$ together with the $\rho$, it would become massless at the same VM fixed point. It would satisfy the same mass formula as the $\rho$ in the form $m_\omega^2=a_\omega f_\pi^2 g^2_\omega$, with $a_\omega\approx a_\rho$ and $g_\omega\approx g_\rho$. This follows from the HLS strategy. This works fairly well in the vacuum and also at low density. But  at high density the $U(2)$ symmetry seems to be badly broken and hence the $\omega$ mass does not seem to go to the VM fixed point arrived at by the $\rho$.

\subsection{Scale symmetry}

The scalar meson of a mass $\sim 600$ MeV denoted in the nuclear literature as $\sigma$ --- and $f_0 (500)$ in the particle-data booklet --- has been a mysterious object since a long time. For  the conundrum associated with this object in nuclear physics, we refer to \cite{conundrum}. Here, we take the point of view adopted in \cite{PKLMR} --- and in previous works referred to therein --- that the scalar could be treated as a dilaton arising from spontaneous breaking of scale symmetry which is  explicitly broken by the trace anomaly of QCD.  We will avoid delving into the long-standing controversy as to whether the association of the scalar with an infrared fixed point makes sense for two or three flavors we are concerned with. It is of course a fundamental issue in theoretical physics. Here we entertain the possibility along the line adopted in the condensed matter circle on emergent symmetries --- and argue for evidence --- for the emergence of scale symmetry in a dense medium even though it may be absent or hidden in QCD in the vacuum. We liken this situation to the hidden gauge symmetry associated with the $\rho$ meson in baryonic system.

The crucial point our argument relies on is that scale symmetry could  actually be present but hidden in QCD~\cite{CCT}. For this argument, we exploit Yamawaki's simple argument developed in connection with walking technicolor theory involving a conformal window~\cite{yamawaki-hiddenSS}. Briefly stated, the argument goes as follows.

Starting with the linear sigma model, by making a series of field redefinitions, one can arrive at the chiral-invariant Lagrangian in terms of the chiral field $U$ and the chiral scalar field $\sigma$ as

\begin{eqnarray}
{\cal L}_{L\sigma}
 &=& \frac{1}{2} \left(\partial_\mu \sigma\right)^2+ \frac{1}{4}{\sigma}^2\cdot {\rm Tr} \left(\partial_\mu U \partial^\mu U^\dagger\right) -\zeta V(\sigma),
 \label{LagM}
\end{eqnarray}
where $V$ is the potential depending on $\sigma$ and $\zeta$ is a constant to be dialled to between zero and $\infty$.

Let us consider two extreme limits: the strong coupling limit $\zeta\to\infty$ and weak coupling limit $\zeta\to 0$.

First  in the strong coupling limit,  $V\to 0$ which gives $\la\sigma\ra\rightarrow f=f_\pi$. Then one simply gets the familiar nonlinear sigma model
\begin{equation}
 {\cal L}_{L\sigma M}  \mathop{\longrightarrow}^{\zeta \rightarrow \infty}  {\cal L}_{NL\sigma}  =
   \frac{f_\pi^2}{4}\cdot {\rm Tr} \left(\partial_\mu U \partial^\mu U^\dagger\right)    \,.
   \label{NL}
    \end{equation}
Note that there is no scale symmetry in this case. The hidden scale symmetry gets shoved into the kinetic energy term.

Now we turn to the weak coupling limit.
Define the scale-dimension-1 and mass-dimension-1 field $\chi$,  called the ``conformal compensator field,"
\be
\chi =f_\chi e^{\sigma/f_\chi}\,.
\ee
Under  scale transformation, $\chi$ transforms linearly while $\sigma$ transforms nonlinearly:
 \begin{equation}
\delta \chi=(1+x^\mu \partial_\mu) \chi\,, \qquad
\delta \sigma= f_\chi+x^\mu \partial_\mu\sigma\,.
 \end{equation}
Here $f_\chi$ is the decay constant. Expressed in terms of the field $\chi$, the Lagrangian (\ref{LagM}) can be written as
\be
  {\cal L}_{L\sigma M} &=& {\cal L}_{\rm sinv} - V(\chi)\,\label{L-dilaton}
\ee
with
\be
{\cal L}_{\rm sinv} &=& \frac{1}{2} \left(\partial_\mu \chi \right)^2+ \frac{f_\pi^2}{4}\left(\frac{{\chi}}{f_\chi}\right)^2\cdot {\rm Tr} \left(\partial_\mu U \partial^\mu U^\dagger\right)\,, \label{s-inv}\\
 V(\chi) &=& \frac{\zeta}{4} f_\chi^4 \left[\left(\left(\frac{\chi}{f_\chi}\right)^2 -1\right)^2-1\right]  \,,\label{potv}
 \ee
with $\frac{\del}{\del\chi} V(\chi)|_{\la\chi\ra=f_\chi}=0$.
The first term of (\ref{L-dilaton}) is scale invariant with scale breaking lodged entirely in the potential (\ref{potv}). It is important to note that scale invariance is obtained in the limit $\zeta\rightarrow 0$ from a linear sigma model.  As is well known, scale symmetry cannot be spontaneously broken if $\zeta$ is exactly zero. This limiting process will figure below in Sec.~\ref{sec:DLFP} as approaching  the ``dilaton-limit fixed point (DLFP)."

The potential (\ref{potv}) is the first term in a more general potential that is anchored on the trace anomaly reflecting the dimensional transmutation, an intrinsic property of QCD,
\begin{equation}
 V(\chi)\Bigg|_{\rm anomaly}= \frac{m_\chi^2 f_\chi^2}{4} \left(\frac{\chi}{f_\chi}\right)^4 \left(\ln \frac{\chi}{f_\chi} -\frac{1}{4}\right)
 \label{anomaly}
\end{equation}
where $\zeta$ is buried in $m_\chi$. This yields $\langle \delta V\rangle = -\langle \theta^\mu_\mu\rangle= m_\chi^2 f_\chi^2\frac{\langle \chi^4\rangle}{f_\chi^4}/4=m_\chi^2 f_\chi^2/4$ and  has a minimum at $\langle \chi\rangle=f_\chi$.

\subsection{$bs$HLS Lagrangian}

\label{bshls}

The hidden symmetries and  topology change discussed above are incorporated in an effective theory that we call $bs$HLS, $b$ standing for baryons brought in as solitons, i.e., skyrmions, $s$ for the scalar dilaton and HLS for the hidden local gauge bosons. The $bs$HLS Lagrangian so constructed is defined at the chiral scale {$\Lambda_{\chi}\approx 4\pi f_\pi \sim 1$} GeV. The scale involved in practical calculations in nuclear physics is much lower. In standard chiral perturbation theory (S$\chi$PT), the cutoff, in practice,  is set  at $\sim (400 - 500)$~MeV, integrating out the vector mesons and the dilaton scalar, and hence involves the nucleons and pions only. In our approach, we put the cutoff denoted $\Lambda_{V}$ slightly above the free-space vector meson mass, so that the explicit degrees of freedom we deal with are the vector mesons and the scalar dilaton in addition to the pions and of course the nucleons. To define the effective Lagrangian to do quantum calculations, we need to fix the ``bare" parameters of the Lagrangian. This we imagine doing at the chiral scale {$\Lambda_{\chi}$} where the vector, axial vector and tensor correlators are matched between the EFT and QCD. This is done with the tree-order terms in the EFT and the operator product expansion in QCD~\cite{HY:WilsonMatch}. This procedure endows the bare\footnote{We will skip the quotation marks from here on. By bare we will always mean the ones with the quotation mark.}  parameters of the EFT Lagrangian with QCD variables, particularly nonperturbative ones such as condensates --- quark, gluon, dilaton, etc., condensate associated with the vacuum. Those condensates are scale dependent, so when brought down to the scale $\Lambda_V$ where they figure in the treatment, they will in principle evolve. But this evolution could be ignorable as is generally assumed in the literature for low energy.  We will do the same in this paper. Now in contrast to standard chiral perturbation approaches, however, we take into account the dependence of the condensates on density since the condensates depend on the ``vacuum" and in a nuclear medium the vacuum changes with the density. The density dependence that results from the matching will be called intrinsic density dependence dubbed as IDD for short.

How the IDD figures in nuclear dynamics depends on the density regime involved. For a reason that will become clear, it is convenient to divide the density regime into two, region I (RI) and region II (RII), the former for $n < n_{1/2}$ and the latter for $n\geq n_{1/2}$ with $n_{1/2}$ being the topology change density. It turns out~\cite{PKLMR} that in RI, the IDD is mostly governed by the dilaton condensate $\la\chi\ra$ which gets locked to the quark condensate $\Sigma\equiv \la\bar{q}q\ra$, so the two scale together. Furthermore due to the scale symmetry incorporated into the $bs$HLS Lagrangian, the bare masses of the hadrons also scale with the dilaton condensate. To a very good approximation, what we call the ``master scaling relation"
\be
\frac{m^\ast_N}{m_N} \approx \frac{m^\ast_\chi}{m_\chi} \approx \frac{m^\ast_V}{m_V} \approx \frac{f^\ast_\pi}{f_\pi} \approx \frac{\langle \chi \rangle^\ast}{\langle \chi \rangle}\equiv \Phi ,
\label{scaling}
\ee
where $V=(\rho, \omega)$, holds in RI.  To the leading order in the counting involved with both scale and chiral symmetries~\cite{li-ma-rho}, the hidden gauge coupling $g_V$ and the dilaton-nucleon coupling $g_{\sigma N}$ do not scale in RI.

It should be stressed that this expression (\ref{scaling}) uses approximate equality.  The reason is that in practical calculations at the scale $\Lambda_V < \Lambda_\chi$,  there can be additional corrections that are most often not big but in some cases are important when fine-tuning is involved as in the case of the equilibrium properties of the nuclear matter ground state. A non-negligible case is when many-body forces enter that are of higher order in the scale-chiral counting when the dilaton and the vector mesons are involved. For instance a short-ranged three-body force which lies above the scale-chiral counting, e.g., involving $\omega$ exchanges, when integrated out, gives rise to an induced density dependence referred to in \cite{PKLMR} as DD$_{\rm induced}$. This is to be distinguished from IDD in the sense that DD$_{\rm induced}$ is inherently Lorentz noninvariant, manifesting the spontaneous breaking of Lorentz symmetry in the vacuum modified by density.\footnote{ A good case where this DD$_{\rm induced}$ enters is the explanation of the long lifetime of C14 where the effect of IDD+DD$_{\rm induced}$ affects the tensor force in such a way to make the Gamow-Teller matrix element nearly vanish at $n\sim n_0$~\cite{c14}. This DD$_{\rm induced}$ could account for the difference in the scaling factor $\Phi$, which at $n\approx n_0$ is $\sim 0.85$ in this C14 dating process whereas it is $\sim 0.75$ for the proton gyromagnetic ratio $\delta g_l^p$ in $^{208}$Pb~\cite{FR}.}  We should stress that it should be feasible within the framework of our approach to actually  calculate  DD$_{\rm induced}$'s in various channels. In this work, we resort to available phenomenological inputs to fine-tune the parameters as needed.

We now turn to region II. Here the topology change at $n_{1/2}$ makes a drastic effect on the EOS. The vanishing of $\overline{\Sigma}$ in the skyrmion crystal can be interpreted in $bs$HLS as that the spectral gap vanishes while the chiral order parameter (i.e., the pion decay constant) remains nonzero. As we will argue below, this makes the effective nucleon mass go to a constant $m_0\neq 0$. We will see also that the dilaton condensate $\la\chi\ra$, on the way to the DLFP described below, goes to a constant $\propto m_0$; hence the dilaton mass $m_\sigma$ as well as the $\omega$ mass $m_\omega$, both scaling as $\Phi (n)$ in RI remain unscaling in RII. In a stark contrast, the $\rho$ mass (and also the $a_1$ mass, i.e., isovector vectors~\cite{generalizedvm1,generalizedvm2}) drops to zero as the $\rho$-nucleon coupling $g_\rho$ goes to zero toward the VM fixed point. It was shown in \cite{PKLMR} that this takes place at $n\gsim 20n_0$, way beyond the range of density relevant to massive compact stars. Since the $\rho$ mass, satisfying  the KSRF formula to all orders of loop corrections~\cite{HY:PR}, goes $\propto$ the $\rho$-nucleon coupling in a dense medium, the $\rho$ mass must drop as the density increases. This differs from the $\omega$ mass because of possible strong breakdown of flavor $U(2)$ symmetry for the vectors $\rho$ and $\omega$ in RII.    It will be seen below that the intricate interplay of the DLFP (explained below) and the VM fixed point leads to an extremely simple pseudoconformal structure in RII.

\section{Dilaton-limit fixed point}

\label{sec:DLFP}

Starting with the parity-doublet $bs$HLS Lagrangian constructed above, we dial a parameter of the EFT Lagrangian to arrive at what is called DLFP theory~\cite{beane-vankolck}. The procedure is tantamount to going from a nonlinear sigma model, hidden-gauge symmetrized, to a linear sigma model in the unraveling of the hidden symmetries in QCD described above.

First we simplify the $bs$HLS Lagrangian by dropping $O(p^4)$ terms. In fact this approximation can be justified. We will treat the $\rho$ meson and $\omega$ meson as the $SU(2)$ and $U(1)$ gauge fields, respectively. In fact, this particular property of the $\omega$ meson is very important with respect to that of the dilaton scalar $\chi$ as the density increases.

\subsection{Dialing to parity doubling}

We start with $bs$HLS with a chiral-invariant nucleon mass $m_0$ introduced in the Lagrangian. Such a mass term does not exist in QCD proper. In fact  we will see below (as we saw with the skyrmion matter) that the parity-doublet symmetry can arise from strong correlations in a dense medium. We will indeed find it to show up at high density. Here we will simply put it in by hand while keeping consistency with symmetries and then expose it by dialing a parameter similarly to what one does for scale symmetry hidden in QCD as discussed above.

As done above, we will restrict ourselves to  the hidden local symmetry $h(x) = SU(2)_V \times U(1)$.
The $bs$HLS Lagrangian to the leading order in scale-chiral counting can be written as~\cite{DLFP-PD}\footnote{To simply the notations, we do not affix $\ast$ to the parameters to indicate IDDs.}
\begin{eqnarray}
{\cal L} &=& {\cal L}_N + {\cal L}_M + {\cal L}_\chi\,,
\label{dlfplag} \\
\mathcal{L}_{N}
&=& \bar{Q}i\gamma^{\mu}D_{\mu}Q - g_{1}f_{\pi}\frac{\chi}{f_{\chi}}\bar{Q}Q
{}+ g_{2}f_{\pi}\frac{\chi}{f_{\chi}}\bar{Q}\rho_{3}Q\nonumber\\
&& - im_{0}\bar{Q}\rho_{2}\gamma_{5}Q
+ g_{v\rho} \bar{Q}\gamma^{\mu}\hat{\alpha}_{\parallel \mu}Q
{}+ g_{v0} \bar{Q}\gamma^{\mu}\mbox{tr}\left[\hat{\alpha}_{\parallel \mu} \right]Q\nonumber\\
&&+ g_{A}\bar{Q} \rho_{3} \gamma^{\mu}\hat{\alpha}_{\perp \mu}\gamma_{5} Q\,,
\label{NchiLargrangian} \\
{\mathcal L}_M
& = & \frac{f_{\pi}^2}{f_{\chi}^2} \chi^2\mbox{tr}\left[ \hat{\alpha}_{\perp\mu}
  \hat{\alpha}_{\perp}^{\mu} \right]
{}+ \frac{a_\rho f_{\pi}^2}{f_{\chi}^2} \chi^2\mbox{tr}\left[ \hat{\alpha}_{\parallel\mu}
  \hat{\alpha}_{\parallel}^{\mu} \right]\nonumber\\
&&{}+ \frac{(a_\omega-a_\rho) f_{\pi}^2}{2f_{\chi}^2} \chi^2\mbox{tr}\left[ \hat{\alpha}_{\parallel\mu} \right]
  \mbox{tr}\left[ \hat{\alpha}_{\parallel}^{\mu} \right] \nonumber\\
 &&{}- \frac{1}{2}\mbox{tr}\left[ \rho_{\mu\nu}\rho^{\mu\nu} \right]
{}- \frac{1}{2}\mbox{tr}\left[ \omega_{\mu\nu}\omega^{\mu\nu} \right]\,,
 \\
{\mathcal L}_\chi
&=& \frac{1}{2}\partial_\mu\chi\cdot\partial^\mu\chi {}-V(\chi).
\end{eqnarray}
Here $V(\chi)$ is the dilaton potential which will be specified later, $Q$ is the nucleon doublet
\begin{eqnarray}
Q & = & \left(
          \begin{array}{c}
            Q_1 \\
            Q_2 \\
          \end{array}
        \right),
\end{eqnarray}
which transforms as $Q \to h(x) Q$, the covariant derivative $D_\mu = \partial_\mu - i V_\mu$, $\rho_i$ are the Pauli matrices acting on the parity doublet, $g_{v0}=\frac 12 (g_{v\omega}-g_{v\rho})$, and $a_\omega, a_\rho$,  $g_A$ and $g_{v\rho,v\omega}$
are all dimensionless parameters.

To move towards a chiral symmetric Gell-Mann-L\'evy (GML)-type linear sigma model, we do the field reparameterizations $\Z=U\chi f_\pi/f_\chi = s+i\vec{\tau}\cdot \vec{\pi}$,  defining the scalar $s$, and write (\ref{dlfplag}) composed of  two parts, one that is regular, ${\cal L}_{\rm reg}$, and the other that is singular, ${\cal L}_{\rm sing}$, as $\mbox{tr}(\Z\Z^\dagger)\equiv\kappa^2 = 2\left( s^2 + \pi^{a\,2}\right) \rightarrow 0$.\footnote{It is worth pointing out that this limiting process is equivalent to dialing $\zeta$ to 0 to go from a nonlinear sigma model to scale-symmetric theory via a linear sigma model as was done with (\ref{LagM}) discussed above.} The singular part that arises solely from the scale invariant part of the original Lagrangian (\ref{dlfplag}) takes the form
\begin{eqnarray}
\mathcal{L}_{\rm sing} &=&
\left( g_{v\rho} -g_A \right) {\cal A} \left( 1/\tr \left[ \Z \Z^{\dagger} \right]\right)\nonumber\\
 &+& \left( \alpha -1\right) {\cal B} \left( 1/\tr \left[ \Z \Z^{\dagger} \right]\right)\,,
\label{sing}
\end{eqnarray}
where $\alpha \equiv f_\pi^2/f_\chi^2$ and
\be
{\mathcal A}
&=&
\frac{ -i }{4} \tr \left( \Z \Z^{\dagger} \right)^{-2} \bar{\psi} \Big[  \tr\left( \sbar{\partial} \left(\Z \Z^{\dagger} \right)  \right)\left\{ \Z, \Z^{\dagger} \right\} \nonumber\\
&&- 2 \tr\left( \Z \Z^{\dagger} \right) \left( \Z \sbar{\partial} \Z^{\dagger} + \Z^{\dagger} \sbar{\partial} \Z \right) \Big] \psi \nonumber \\
&& \frac{ -i }{2} \tr\left( \Z \Z^{\dagger} \right)^{-1} \bar{ \psi } \rho_{3} \gamma_{5} \left( \Z \sbar{\partial} \Z^{\dagger} - \Z^{\dagger} \sbar{\partial} \Z \right) \psi \\
{\mathcal B} &=& \frac{-1}{16 \alpha } \mbox{tr} \left( \Z \Z^\dagger \right)^{-1} \mbox{tr} \left[ \partial_\mu \left( \Z \Z^\dagger \right)\right] \mbox{tr} \left[ \partial^\mu \left( \Z \Z^\dagger \right)\right] \,,
\ee
where
\begin{eqnarray}
\psi & = &  \frac{1}{2}\left[ \left( \xi_R^{\dagger}+ \xi_L^{\dagger} \right)
+ \rho_{3} \gamma_5 \left( \xi_R^{\dagger} - \xi_L^{\dagger}  \right) \right] Q.
\end{eqnarray}
That ${\mathcal L}_{\rm sing}$ be absent leads to the conditions that
\be
g_{v\rho}-g_A\rightarrow 0\,,
\quad
\alpha -1 \to 0\,.
\ee
The second condition is precisely the locking of $f_\pi $ and $f_\chi$ mentioned above.
Using large $N_c$ sum-rule and renormalization-group arguments~\cite{beane-vankolck}, we infer
\be
g_A-1\rightarrow 0\,.
\ee
In the density regime where the GML-type linear sigma model is valid, the nucleon mass can be given as
\begin{equation}
m_{N_\pm} = \mp g_2 \langle s \rangle + \sqrt{\left( g_1 \langle s \rangle \right)^2 +  m_0^2}\,,\label{nmass}
\end{equation} where $\langle s \rangle$ is the vacuum expectation value of $s$.
As the chiral symmetry restoration point is approached, $\langle s \rangle\rightarrow 0$, so in the limit $\mbox{tr}(\Z \Z^\dagger) \rightarrow 0$,
we expect
\begin{equation}
m_{N_\pm} \rightarrow m_0\,.
\end{equation}
These are the constraints that lead to the dilaton limit as announced above. It follows then that
\be
g_{\rho NN}=g_\rho(g_{v\rho}-1)\rightarrow 0.
\ee
We thus find that in the dilaton limit, the $\rho$ meson decouples from the nucleon.\footnote{ Note as mentioned above that this decoupling occurs even if the VM where $g_\rho\to 0$ is not reached.} In contrast, the limiting $\mbox{tr}(\Z \Z^\dagger)\rightarrow 0$ {\it does not} give any constraint on $(g_{v\omega}-1)$. The $\omega$-nucleon coupling remains nonvanishing in the Lagrangian which in  unitary gauge  and in terms of fluctuations $\tilde{s}$ and $\tilde{\pi}$
around their expectation values, takes the form
\begin{eqnarray}
{\mathcal L}_N
& = &
\bar{N}i\sbar{\partial}{ N} - \bar{ N}\hat{M}{ N}
{}- g_1\bar{ N}\left(
\hat{G}\tilde{s} + \rho_3\gamma_5 i\vec{\tau}\cdot\vec{\tilde{\pi}}
\right)  N
\nonumber\\
&&{}+ g_2\bar{ N}\left(
\rho_3 \tilde{s} + \hat{G}\gamma_5 i\vec{\tau}\cdot\vec{\tilde{\pi}}
\right) {N} \nonumber\\
& &{}
+ \left(1-g_{v\omega} \right) g_\omega{ N}  \frac{\sbar{\omega}}{2}  {N}\,,
\end{eqnarray}
where $N$ is in a parity eigenstate. This Lagrangian is the same as the one given in \cite{detar-kunihiro} except for the $\omega$-nucleon interaction. This is just the nucleon part of the linear sigma model in which the $\omega$ is minimally coupled to the nucleon, applicable infinitesimally below the critical density $n_c$ with the effective nucleon mass replacing $m_0$.

\subsection{Emergent parity doubling}\label{paritydoubling}

Here we show that parity doubling arises by nuclear correlations from $bs$HLS without parity-doublet symmetry incorporated {\it ab initio} in contrast to what was obtained above by {\it dialing parameters} from the $bs$HLS Lagrangian with the chiral-invariant mass $m_0$ put in by hand~\cite{interplay}. We can do this in the mean-field approximation using the simplified $bs$HLS Lagrangian which is obtained from (\ref{NchiLargrangian}) by turning off $m_0$ and put in parity eigenstates,
\begin{eqnarray}
\mathcal{L}_{N}
&=& \bar{N}i\gamma^{\mu}D_{\mu}N - hf_{\pi}\frac{\chi}{f_{\chi}}\bar{N}N + g_{v\rho} \bar{N}\gamma^{\mu}\hat{\alpha}_{\parallel \mu}N\nonumber\\
&&
{}+ g_{v0} \bar{N}\gamma^{\mu}\mbox{tr}\left[\hat{\alpha}_{\parallel \mu} \right]N
{}+ g_{A}\bar{N}  \gamma^{\mu}\hat{\alpha}_{\perp \mu}
\gamma_{5} N\,,
\label{NchiLargrangian1}
\end{eqnarray}
and $V(\chi)$ is the dilaton potential that we take the form \eqref{potv}.

We consider the EFT Lagrangian effective in a vacuum affected by density. Its bare parameters appropriate in that modified vacuum carry the density dependence via the correlators. This is the IDD plus  DD$_{\rm induced}$ as explained. We will loosely refer to this ``effective" density dependence as IDD$^\ast$ represented in the bare parameters as $\ast$-ed quantities.\footnote{ In working with the thermodynamic potential it is important to properly treat the density dependence of bare parameters as discussed in \cite{song}. Otherwise one loses the rearrangement terms and hence fails to conserve  the energy-momentum tensor.}

The point made in \cite{MLPR} --- and even before in \cite{PKLMR} --- is that the $bs$HLS Lagrangian suitably endowed with the IDD$^\ast$, when treated in the mean field, effectively gives the Landau Fermi-liquid fixed-point structure~\cite{song}.

\begin{figure}[h]
\begin{center}
\includegraphics[width=8cm]{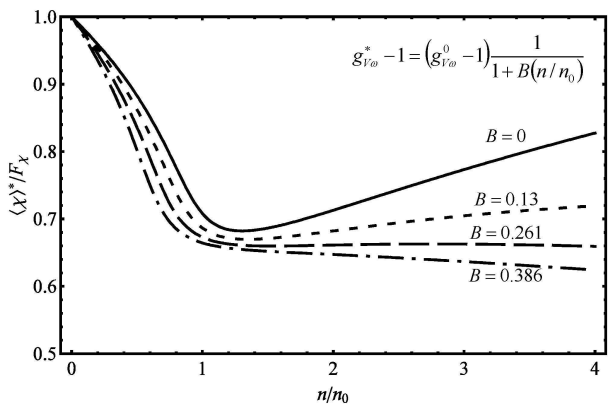}
\caption{The ratio $m_N^*/m_N\approx \la\chi\ra^*/\la\chi\ra_0$  as a function of
density for varying density dependence of $g_{v\omega}^\ast$.
Note that for a given $\omega$-nuclear coupling, the nucleon mass stops dropping at a density $n_A$
above nuclear matter density $n_0$ and stays  constant above that
density. {Figure borrowed from Ref.~\cite{interplay}}.
}
\label{mass_coupling}
\end{center}
\end{figure}

The thermodynamic potential in the mean-field approximation takes the form
\begin{eqnarray}
\Omega(\chi,\, n)&=&
\frac{1}{4\pi^{2}} \left[ 2 E_{F}^{3} p_{F} - m_{N}^{\ast 2} E_{F} p_{F}
{}- m_{N}^{\ast 4} \ln \left( \frac{E_{F} + p_{F} }
{m_{N}^{\ast}} \right) \right]\nonumber\\
&&{} + \frac{\left(g_{v\omega}^\ast -1 \right)^2}
{2a_\omega f_{\pi}^2
{\chi}^2/f_{\chi}^2} n^2
{} -  V(\chi) -\mu(n) n\,,
\label{omega}
\end{eqnarray}
where $E_F = \sqrt{p_F^2 + m_N^{\ast\, 2}}$
and the chemical potential is given as a function of density $n$ by
\begin{eqnarray}
\mu(n) & = & E_F(n) \nonumber\\
& &{} + \frac{\left( g_{v\omega}^\ast -1\right)^2}
{a_\omega f_{\pi}^2{\chi}^2/ f_\chi^2} n
{}+ \frac{\left( g_{v\omega}^\ast -1\right)}
{a_\omega f_{\pi}^2{\chi}^2/ f_\chi^2}n^2
\frac{\partial \left( g_{v\omega}^\ast -1\right) }{\partial n}\,.
\nonumber\\
\end{eqnarray}
The nucleon mass is connected to the $\omega$-nucleon coupling by the
equation of the motion for $\chi$ and $\omega$, and the in-medium property
of the $\chi$ condensate --- equivalently the in-medium mass of the dilaton
--- controls
the behavior of the nucleon mass at high density. The nucleon mass depends
on $\bchi=\la \chi\ra$ via
\be
m_N^\ast = h\bchi\,.
\ee
The gap equation for $\chi$ is
\be
&& \left[
\frac{ m_N^2}{\pi^2 f_\chi^2}\left(
p_F E_F - m_N^{\ast\, 2}\ln\left(\frac{p_F + E_F}{m_N^\ast}\right)
\right) \right. \nonumber\\
&&\left.\quad {}- \frac{\left(g_{v\omega}^\ast -1 \right)^2}
{a_\omega f_{\pi}^2 \chi^4/f_{\chi}^2} n^2
 {}+ \frac{ m_\chi^2}{2} \left( \frac{\chi^2}{f_\chi^2} \right)
\ln\left(\frac{\chi^2}{f_\chi^2}\right)\right] \chi = 0\,.\nonumber\\
\label{gapchi}
\ee
In the mean-field approach, the dilaton limit is reached as $\bchi \rightarrow 0$.  Suppose
the $\omega$-nucleon coupling drops slowly. This not only causes the nucleon
mass to drop slowly, but also delays the dilaton limit, $g_{A} = g_{v\rho}=1$, to higher density. This feature can be seen in Fig.~\ref{mass_coupling} given in  Ref.~\cite{interplay}.
Let us take the scaling of the $\omega$-nucleon coupling in the simple form
\begin{equation}
\frac{g_{v\omega}^\ast - 1}{g_{v\omega}-1}
= \frac{1}{1 + B n/n_0}\,. \label{omega_coupling}
\end{equation}
Here the scaling of the hidden gauge coupling $g_\omega$ is ignored, which is negligible. Thus only the scaling of the effective coupling $g_{v\omega}$ intervenes.

For a given constant $B$, the nucleon mass is calculated by fitting the
binding energy and the pressure of nuclear matter at $n_0$. The two density-dependent quantities involved are $m_\chi^*$ and
$g_{\omega}^*$ that are determined by the binding energy and the pressure
at $n=n_0$ for given $B$.  The result is plotted in Fig.~\ref{mass_coupling}. Remarkably  the nucleon mass is found to drop almost linearly in density to about 70\% of the free-space mass at a density denoted $n_A$ above $ n_0$. Up to $\sim n_0$, the dilaton condensate, locked to the quark condensate,  is consistent with the empirical value of the quark condensate estimated from the in-medium pion decay constant measured in deeply bound pionic states~\cite{PionBound}. It then stabilizes to a constant for $n\gsim n_A$. We identify this density {$n_A$} with the skyrmion-to-half-skyrmion transition density $n_{1/2}$ we encountered above.

How this comes about is an intricate interplay between the nucleon mass and the $\omega$-$NN$ coupling after $n\sim n_A$. This is explained in the Appendix.
Here we should stress that this mean-field calculation was made with  $m_0=0$. Nevertheless, we have found
$m_N^\ast \sim 0.7 m_N$ in high density, indicating that a nonvanishing $m_0$
emerges dynamically. The interplay between the nucleon mass and the $\omega$-nucleon coupling
as revealed in this way is  similar to what was found by the renormalization group equation analysis~\cite{interplay}  and consistent with what was phenomenologically
observed in nuclear EFT description with IDD$^\ast$  modified by the topology
change~\cite{PKLMR}.

In brief, this analysis suggests that as density reaches $n_A\sim n_{1/2}$ the nucleon mass goes as
\be
m_N\propto \la\chi\ra\sim {\rm const},
\ee
and parity doubling emerges via an interplay between $\omega$-nuclear coupling and the dilaton condensate.

\section{Pseudoconformal equation of state}

\label{sec:PCM}

\subsection{Trace of energy-momentum tensor at $n\geq n_{1/2}$}

We now argue that the emergence of parity doubling, scale symmetry and hidden local symmetry at $n\geq n_{1/2}$ has a drastic impact on the star structure.  For this, we look at the trace of  the energy-momentum tensor. First we compute the TEMT in the mean-field approximation with the $bs$HLS Lagrangian. From the Lagrangian Eq.~(\ref{NchiLargrangian1}) together with the dilaton potential \eqref{potv}, one obtains the energy density $\epsilon$ and the pressure $P$ (at $T=0$) of the form
\begin{eqnarray}
\epsilon
&=& \frac{1}{4\pi^2} \left[ 2E_F^3 k_F - m_N^{\ast\,2}E_F k_F - m_N^{\ast\,4} \ln\left( \frac{E_F + k_F}{m_N^\ast}\right) \right]  \nonumber\\
&&{} + g_\omega \left( g_{v\omega} -1 \right) \langle \omega_0 \rangle n -\frac{1}{2} a_\omega  f_{\pi}^2 g_\omega^2 \frac{\langle \chi \rangle^2}{f_\chi^2} \langle \omega_0 \rangle^2 +  V(\langle \chi \rangle) \nonumber\\
\label{eden}
\end{eqnarray}
and
\begin{eqnarray}
P &=& \frac{1}{4\pi^2} \left[ \frac{2}{3}E_F k_F^3 - m_N^{\ast\,2}E_F k_F + m_N^{\ast\,4} \ln\left( \frac{E_F + k_F}{m_N^\ast}\right) \right] \nonumber\\
&& + \frac{1}{2} a_\omega f_{\pi}^2 g_\omega^2 \frac{\langle \chi \rangle^2}{f_\chi^2} \langle \omega_0 \rangle^2 -  V(\langle \chi \rangle)\,.\label{pre}
\end{eqnarray}
Using the solutions of the gap equations for $\chi$ and $\omega$ that follow from extremizing (\ref{omega}), i.e.,
\begin{eqnarray}
&\frac{m_N^2\langle \chi \rangle}{\pi^2 f_\chi^2} \left[ k_F E_F - m_N^{\ast\,2} \ln \left( \frac{k_F + E_F}{m_N^\ast} \right) \right] -\frac{a_\omega f_{\pi}^2}{f_\chi^2} g_\omega^2 \langle \omega_0\rangle^2 \langle \chi \rangle\nonumber\\
& + \left. \frac{\partial\, V(\chi)}{\partial \chi} \right|_{\chi = \langle \chi \rangle} =0\,,
 \label{gap1}\\
&g_\omega \left(g_{v\omega}-1 \right)n -a_\omega f_{\pi}^2 g_\omega^2 \frac{\langle \chi \rangle^2}{f_\chi^2} \langle \omega_0 \rangle = 0\,, \label{gap2}
\end{eqnarray}
it is straightforward to derive from (\ref{eden}) and (\ref{pre}) the vacuum expectation value (VEV) of the TEMT $\theta_\mu^\mu$ (we work in the chiral limit)
\begin{eqnarray}
\langle \theta^\mu_\mu \rangle
&=& \langle \theta^{00} \rangle - \sum_i \langle \theta^{ii}\ra = \epsilon - 3 P\nonumber\\
& =& 4V(\langle \chi \rangle) - \langle \chi \rangle \left. \frac{\partial V( \chi)}{\partial \chi} \right|_{\chi = \langle \chi \rangle}. \label{TEMT1}
\ee
What is significant of this result is that in the mean field of $bs$HLS, the TEMT is given solely by the dilaton condensate. This is in the chiral limit, but we expect this relation to more or less hold for a small pion mass. From what we learned from above, i.e.,that the emergence of parity doubling at $n\gsim n_{1/2}$ implies $\la\chi\ra\to cm_0$ where $c$ is a constant,  we have
\be
\langle \theta^\mu_\mu \rangle\propto f(m_0)\neq 0\ \ {\rm for}\ \ n\gsim n_{1/2}. \label{constantTEMT}
\ee

As stated~\cite{PKLMR,MLPR}, the mean-field treatment of $bs$HLS amounts to doing Landau Fermi-liquid fixed-point approach ignoring corrections of $O(1/\bar{N})$, where $\bar{N}=k_F/(\Lambda - k_F)$ with $\Lambda$ being the cutoff above the Fermi sea. In \cite{PKLMR}, the corrections to the Fermi-liquid fixed-point approximation were included in the so-called ``$V_{lowk}$ RG" formalism. \footnote{The $V_{lowk}$ renormalization group (RG) employed in~\cite{PKLMR}, very well known in nuclear theory community, was explained in detail there. There the formalism reviewed in \cite{vlowk} was updated so as to incorporate the structure of $bs$HLS.  Briefly for those outside of the field, the $V_{lowk}$RG purports to do a Wilsonian renormalization group effective field theory treatment of many-nucleon systems that go beyond the Fermi-liquid fixed-point approximation.}  It was found that with $n_{1/2}=2n_0$, the TEMT satisfied the behavior (\ref{constantTEMT}) for both nuclear matter ($\alpha=0$) and pure neutron matter ($\alpha=1$) and hence for $\beta$-equilibrated compact stars.

\subsection {Pseudoconformal model (PCM)}
As argued in detail in \cite{interplay} and recounted briefly above, the parity-doubling approaching the dilaton-limit fixed point arises due two crucial effects taking place in the $n\geq n_{1/2}$ regime. One is the $\rho$ meson decoupling from the system and the other the interplay in the $\omega$ coupling to nucleons.

 It is mysterious that these effects lead to a function uniquely of the density-independent quantity $m_0$ in the TEMT for $n\geq n_{1/2}$. We do not have an understanding of how this comes about. What is robust is that it leads to the sound speed of stars $v_s^2/c^2=1/3$,  usually associated with conformal symmetry with a vanishing energy-momentum tensor, hence called ``conformal  sound velocity." What is relevant here is that it also arises when the TEMT is density independent. This is easily seen from that
 \be
\frac{\partial}{\partial n} \la\theta_\mu^\mu\ra=\frac{\partial \epsilon (n)}{\partial n} (1-3{v_s^2})=0\label{derivTEMT}
\ee
with $v_s^2=\frac{\partial P(n)}{\partial n}/\frac{\partial \epsilon(n)}{\partial n}$.  Since $\frac{\partial \epsilon (n)}{\partial n} \neq 0$ in the range of densities involved, we immediately obtain the sound velocity $v_s^2/c^2 = 1/3$. Since the TEMT is not equal to zero, it is appropriate to call it  ``pseudoconformal sound velocity."

In \cite{PKLMR,MLPR}, it was shown that this pseudoconformal property could simply be captured by a two-parameter formula. Consider
 $E/A$ for $n\geq n_{1/2}$ in the form
\be
E/A= - m_N +X^\alpha x^b + Y^\alpha x^d\ \ {\rm with}\ x\equiv n/n_0
\ee
where $X$, $Y$, $b$ and $d$ are parameters to be fixed and $\alpha=(N-Z)/(N+Z)$. The sound velocity takes the form
\be
v_s^2=\frac{dP/dx}{d\epsilon/dx}=\frac{X^\alpha b(b+1)x^b+Y^\alpha d(d+1)x^d}{X(b+1)+Y(d+1)x^d},
\ee
where $P$ is the pressure and $\epsilon$ is the energy density. If we choose $d=-1$ and $b=1/3$, then the  $E/A$ given by
\be
E/A= - m_N +X^\alpha  x^{1/3} + Y^\alpha x^{-1}\ \ {\rm with}\ x\equiv n/n_0 \nonumber\\
\label{PC-RII}
\ee
has the sound velocity
\be
v_s^2/c^2=1/3\label{pc-sound}
\ee
independently of $X^\alpha$ and $Y^\alpha$.

The PCM for the EOS provides then $E/A$ given by the union of that given by $V_{lowk}$ in RI ($ n<n_{1/2}$) --- that is constrained by the properties of normal nuclear and slightly above reached by experiments --- and that given by (\ref{PC-RII}) in RII ( $n\geq n_{12}$) --- that embodies pseudoconformal structure --- with the parameters $X^\alpha$ and $Y^\alpha$ fixed by the continuity at $n=n_{1/2}$ of the chemical potential and pressure
\be
\mu_{\rm I}=\mu_{\rm II},\ P_{\rm I}=P_{\rm II}\ \ {\rm at} \ \ n=n_{1/2}.
\label{eq:FixCondition}
\ee

\section{Compact Stars}
\label{sec:star}

We now apply the PCM formulated  above to set both the lower and upper bounds of the topology changeover densities  from the structure of dense compact-star matter. The lower bound is indicated by the recent gravitational wave data on the tidal deformability and the upper bound by the available heavy-ion experimental data.

\subsection{Tidal deformability and the lower bound of $n_{1/2} $}

It was shown in \cite{PKLMR} that the property of the trace of the energy-momentum tensor in RII ($n\geq n_{1/2}$) going as $\la\theta_\mu^\mu\ra\propto \la\chi\ra^4$ with the condensate becoming $\la\chi\ra$ density independent, calculated in the mean field in $bs$HLS, was exactly reproduced by the $V_{lowk}$RG calculation. This equality was verified when the topology change was taken at $n_{1/2}=2n_0$. While the mean-field calculation relies on the Fermi-liquid fixed-point approximation that ignores $O(1/\bar{N})$ corrections, the $V_{lowk}$RG calculation includes (in principle) all orders of $1/\bar{N}$ in the ring-diagram approach~\cite{ringdiagram1,ringdiagram2}. To the extent that the ring-diagram approach which goes beyond the Fermi-liquid fixed point is reliable for the many-body problem at near the equilibrium density $n_0$, one expects it to remain valid before the topology change takes place. Assuming that the validity holds up to $n_{1/2}$, we can take the result of \cite{PKLMR} as a support for the PCM EOS for the case of $n_{1/2}=2n_0$. Since the $V_{lowk}$RG approach successfully explains {\it all} properties of symmetric nuclear matter at $n=n_0$ and even up to near $2n_0$ as measured in heavy-ion experiments (e.g., the symmetry energy at $n=2n_0$~\cite{bal}; see below), that the PCM --- that treats RI in $V_{lowk}$ and RII with (\ref{PC-RII}) --- reproduces the full $V_{lowk}$RG results supports the intricate scaling behavior in RII being captured by the pseudoconformal structure.

There was, however, one potentially significant problem in the calculation of \cite{PKLMR} that was revealed by the recent bound established on the dimensionless tidal deformability $\Lambda < 800$~\cite{GW170817} for a $1.4$-solar-mass ($M_{1.4}$) star. While overall star properties are fairly well explained --- and that includes the $\gsim 2$-solar mass and its radius --- the $\Lambda$ predicted by the theory came out to $\sim 790$~\cite{PKLMR}, which is a bit too high, given that a more refined analysis of gravitational waves seems to point to a lower bound~\cite{abottetal} than what was announced in \cite{GW170817}. Why this result can pose a problem for the PCM with $n_{1/2}=2n_0$ is that the central density of the $M_{1.4}$ star comes out to be $n\gsim 2n_0$ (see Table \ref{tab:StarProperty}) in the PCM, hence coincident with the transition density where the EOS goes from ``soft" as needed at $n\sim n_0$ to ``hard"  above $n_0$ so as to accommodate the maximum mass $\gsim 2.02 M_\odot$. This means that the changeover density must be {\it higher} than $2n_0$, thus setting the lower bound for the topology change density,
\be
n_{1/2} > 2n_0.
\ee
It has indeed been verified that a higher $n_{1/2}$ could resolve this problem. In \cite{MLPR}, the PCM with $n_{1/2}=2.6n_0$ is found to give,  for $M_{1.4}$, $\Lambda\approx 640$ with the central density at $n_c\approx 2.3n_0$, which is in the soft region RI. The resulting star properties, however, remain practically the same as the case of $n_{1/2}=2n_0$. This consolidates the observation made before --- and reconfirmed below --- that star properties are fairly insensitive to the location of $n_{1/2}$ in the vicinity of $\sim 2n_0$, where quark degrees of freedom are expected to figure.

The question then is, how far can one  increase $n_{1/2}$ without upsetting the good star properties? In particular we are interested in how the range of density allowed by the location of $n_{1/2}$ compares with the range of the  baryon-quark continuity as in the phenomenological model of \cite{baymetal}. This question is highly relevant to the possible applicability of the notion of Cheshire cat to dense matter.

\subsection{Analysis for  $2 \leq (n_{1/2}/n_0) \leq 4$}

{
Now, we are in a position to make an explicit calculation to see the impact of the topology change at $n_{1/2}$  on the equilibrium  nuclear matter as well as the star properties.

As explained in \cite{PKLMR}, the scaling parameters may be minimally fine-tuned in RI\footnote{ It is well recognized that the ground-state properties of the equilibrium nuclear matter are extremely sensitive to parameters. In the $bs$HLS framework, this can be understood as differences in the cutoff scales involved with DD$_{\rm induced}$ for different mesonic degrees of freedom, i.e., $\rho$, $\omega$, $\chi$, $N$, etc.} to fit the nuclear matter properties around the saturation density and the results for the $V_{lowk}$ calculation in RI can be fit extremely well  by the simple function
\begin{eqnarray}
E/A & =
& A^\alpha \left(\frac{n}{n_0}\right) + B^\alpha\left(\frac{n}{n_0}\right)^{D^\alpha}
\label{eq:FitingI}
\end{eqnarray}
with $\alpha = (N-Z)/(N+Z)$. We obtain to a high precision the parameters of the fitting function ~\eqref{eq:FitingI} in RI. They are listed in Table~\ref{tab:FitRI}.   As expected,  the parameters in the fitting function vary a bit depending on the topology change density $n_{1/2}$ due to the fine-tuning needed for the properties of normal nuclear matter.
\begin{table}[!h]
\centering
\caption{Fitting parameters (in of MeV) in RI for symmetric nuclear matter $(\alpha=0)$ and neutron matter $(\alpha = 1)$ with different choices of $n_{1/2}$.}
\label{tab:FitRI}
\begin{tabular}{c|cc|cc|cc}
\hline
\hline
\multirow{2}{*}{$n_{1/2}/n_0$}~& \multicolumn{2}{c|}{$A_I$}&\multicolumn{2}{|c}{ $B_I$}&\multicolumn{2}{|c}{ $D_I$}\cr\cline{2-7}
&~$\alpha=0$~&~$\alpha=1$~&~$\alpha=0$~&~$\alpha=1$~&~$\alpha=0$~&~$\alpha=1$
\cr
\hline
2.0~\cite{PKLMR} &${}-45.5$&9.11&30.1&2.14&1.54&4.08\cr\hline
$3.0$ &${}-27.0$&6.88&11.4&4.68&2.12&2.87\cr\hline
$4.0$ &${}-24.0$&4.09&8.86&6.84&2.28&2.57\cr
\hline
\hline
\end{tabular}
\end{table}

Now in RII, $E/A$ is simply given by ~\eqref{PC-RII} with the continuity condition~\eqref{eq:FixCondition}. The parameters of the function~\eqref{PC-RII} are summarized in Table~\ref{tab:FitRII}. We note that the fitting parameters in RII are strongly dependent on the position of $n_{1/2}$, particularly for the pure neutron matter and $n_{1/2}=4n_0$.
\begin{table}[!h]
\centering
\caption{Fitting parameters (in unit MeV) in RII for symmetric nuclear matter $(\alpha=0)$ and neutron matter $(\alpha = 1)$ with different choice of $n_{1/2}/n_0$.}
\label{tab:FitRII}
\begin{tabular}{c|cc|cc}
\hline
\hline
\multirow{2}{*}{$n_{1/2}/n_0$}~& \multicolumn{2}{c|}{$X^\alpha$}&\multicolumn{2}{|c}{ $Y^\alpha$}\cr\cline{2-5}
&~$\alpha=0$~&~$\alpha=1$~&~$\alpha=0$~&~$\alpha=1$~
\cr
\hline
$2.0$~\cite{PKLMR} &{}570&686&440&253\cr\hline
$3.0$ &{}575&725&423&64.4\cr\hline
$4.0$ &{}607&912&247&${}-946$\cr
\hline
\hline
\end{tabular}
\end{table}

In Fig.~\ref{fig:CompEnergy} is given the $E/A$ predicted with the parameters listed in Tables~\ref{tab:FitRI} and~\ref{tab:FitRII}. It is noteworthy that as the density exceeds $n_{1/2}$ with the appearance of the pseudoconformal structure, the energy of the system $E/A$ gets greater the higher $n_{1/2}$ is. In contrast, the property of the ground state of nuclear matter is unaffected by the location of $n_{1/2}$: At the saturation density $n_0 \simeq 0.16~{\rm fm}^{-3}$, the binding energy $BE$ and the compression modulus $K$ are given, for all $n_{1/2}$ considered, by
\begin{eqnarray}
BE & = & {}  15.65~{\rm MeV}; \quad K = 228.9~{\rm MeV},
\end{eqnarray}
respectively, in consistency with the values widely quoted in the literature.
\begin{figure}[!h]
 \begin{center}
   \includegraphics[width=8cm]{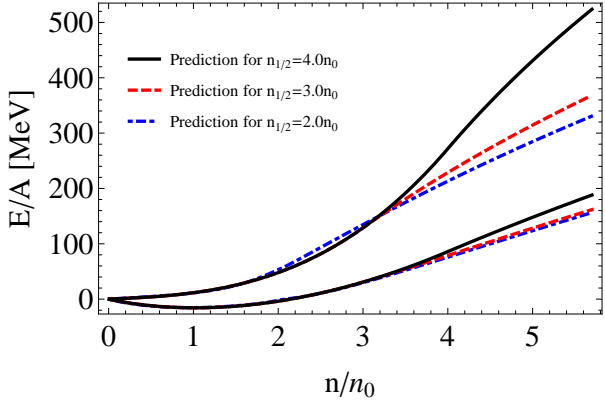}
  \end{center}
 \caption{Predicted $E/A$ vs density for $n_{1/2}/n_0=2.0$, $3.0$ and $4.0$. The upper (lower) curves are for pure neutron matter with $\alpha=1$ (symmetric nuclear matter with $\alpha=0$)}.
\label{fig:CompEnergy}
\end{figure}

The symmetry energy $E_{sym}$ predicted by the theory is given in  Fig.~\ref{fig:Esym}.  As shown in \cite{PKLMR} in the $V_{lowk}$RG formalism, the cusp found at the quasiclassical approximation~\cite{cusp} is smoothed by higher-order $1/\bar{N}$ corrections. But it clearly reflects the changeover from soft to hard in the EOS at $n_{1/2}$: The greater $n_{1/2}$, the harder the $E_{sym}$ becomes.  Up to $n\sim 2n_0$,  the symmetry energy is insensitive to $n_{1/2}$. 
 It is consistent with the available empirical constraints. The predictions
\begin{eqnarray}
E_{sym}(n_0) & =
& 27.2~{\rm MeV}, \;\; E_{sym}(2n_0) = 51.7~{\rm MeV}
\label{eq:Esymn0}
\end{eqnarray}
more or less agree with the constraints $E_{sym}(n_0) = 31.7 \pm 3.2$~MeV and $E_{sym}(2n_0) = 46.9 \pm 10.1$~MeV  obtained recently~\cite{bal}.

\begin{figure}[h]
 \begin{center}
   \includegraphics[width=8cm]{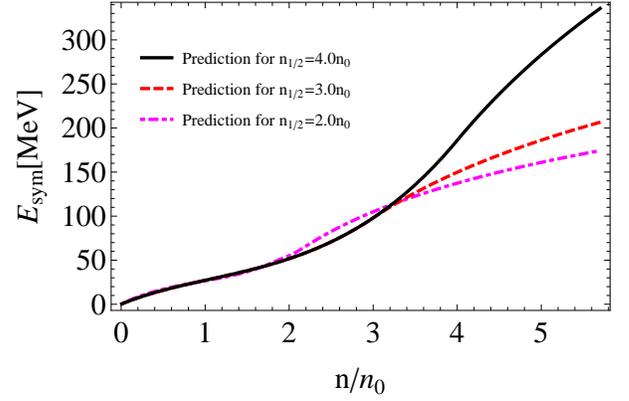}
  \end{center}
 \caption{Predicted $E_{sym}$ vs density.}
\label{fig:Esym}
\end{figure}

We now turn to the EOS that enters into the Tolman-Oppenheimer-Volkoff (TOV) equation.

Fully equipped with the energy density $\epsilon (n)$ gotten via $E/A$ calculated above
\begin{eqnarray}
\epsilon(n) & =
& n \left[\frac{E_0(n)}{A} + m_N\right],
\label{eq:edn}
\ee
and the pressure density
\be
p(n) & =
& n \frac{d \epsilon(n)}{dn} - \epsilon(n),
\label{eq:Pn}
\ee
we are now ready to proceed to predict the star properties.

First,  the pressure density for neutron matter as a function of density for the given locations of $n_{1/2}$ is given  in Fig.~\ref{fig:Press}.  The results are compared with the bounds presently available from heavy-ion experiments~\cite{Tsang-Press}.
\begin{figure}[h]
 \begin{center}
   \includegraphics[width=8cm]{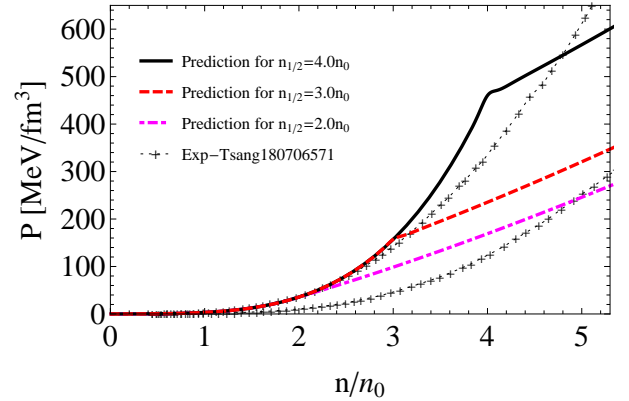}
  \end{center}
 \caption{Predicted pressure  vs density compared  with the available experimental bound given by~\cite{Tsang-Press}.}
\label{fig:Press}
\end{figure}
While the pressures obtained for $n_{1/2}/n_0=2\ {\rm and}\  3$ are found fully consistent with the bounds up to $n=5n_0$,  that for $n_{1/2}=4.0 n_0$ deviates from the experimental bounds above $\sim 3n_0$.  This result suggests the upper bound for $n_{1/2}$,
\be
n_{1/2} < 4n_0.
\ee

Given the EOS described above, we are now in position to fully analyze the compact-star properties using the TOV equation. We follow the same procedure as in~\cite{PKLMR}.  As there, we take into account the presence of leptons in beta equilibrium in solving the TOV equation.

We first plot  in Fig.~\ref{fig:MvsR} the mass-radius relation of the compact stars for different choice of $n_{1/2}$.
\begin{figure}[h]
 \begin{center}
   \includegraphics[width=8cm]{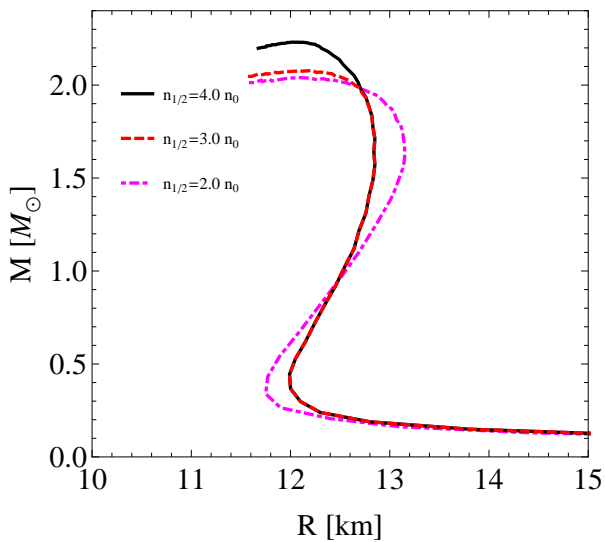}
  \end{center}
 \caption{Mass-radius relation of compact stars with different choice of $n_{1/2}$. Note that below $M\approx 2M_\odot$, the curves for $n_{1/2}/n_0 =3.0\ {\rm and} \ 4.0$ represented in red with black dots are coincident.}
\label{fig:MvsR}
\end{figure}
It is found that the higher the $n_{1/2}$, the larger the upper bound of the star masses.  The upper bound comes out to be roughly  $(2.04M_\odot {\rm -} 2.23 M_\odot)$ for $2.0 \leq n_{1/2}/n_0 \leq 4.0$. This bound is consistent with the observation of the massive neutron stars~\cite{Nat-2solar,Sci-2solar}. It is notable that, when $n_{1/2} \geq 3.0 n_0$, changing the position of $n_{1/2}$  affects only the compact stars with mass $\lesssim 2.0 M_\odot$ although the mass-radius relation is affected by the topology change when $2.0 n_0 \leq n_{1/2} \leq 3.0 n_0$.

In Fig.~\ref{fig:MvsNc} is plotted the star mass vs. the central density of the stars.  What is noteworthy is that the maximum central density of the stars is about  $\sim (4{\rm -}5)n_0$, more or less independent of the topology change density.
\begin{figure}[h]
 \begin{center}
   \includegraphics[width=8cm]{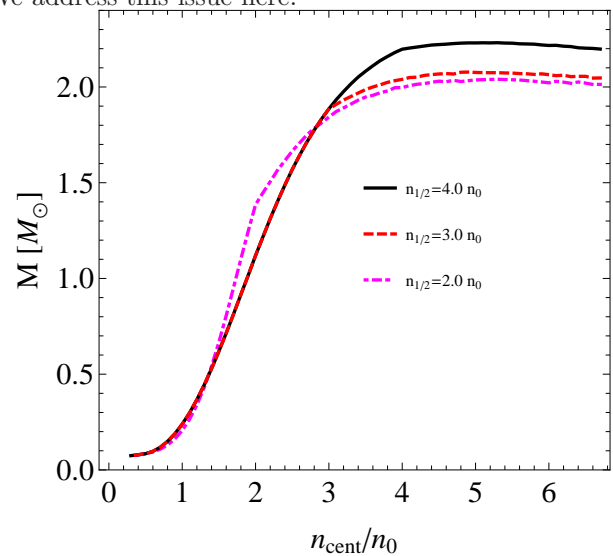}
  \end{center}
 \caption{The star mass vs the central density. The red curve with black dots stands for $n_{1/2}/n_0= 3\ {\rm and}\ 4$ }
\label{fig:MvsNc}
\end{figure}

We now turn to the star properties that distinguish the pseudoconformal model from all others found in the literature, namely, the sound velocity $v_s^2/c^2=1/3$ and the tidal deformability $\Lambda$.

First the sound velocity.

The PCM, derived from the result of \cite{PKLMR} obtained in the {\it full} $V_{lowk}$RG formalism with $bs$HLS for $n_{1/2}=2n_0$ and confirmed in \cite{MLPR} for $n_{1/2}=2.6 n_0$, when applied to $n_{1/2}> 2.0 n_0$, is by construction to  yield the conformal sound velocity $v_s^2/c^2=1/3$ for density $\gsim n_{1/2}$. As stated, this can be taken as representing the signal for a change of degrees of freedom from baryons to QCD degrees of freedom at that density. We will discuss below what this means in terms of observability of this prediction and the role of hidden symmetries of QCD.

Next the tidal deformability $\Lambda$ in gravitational waves.

As discussed in \cite{MLPR}, the tidal deformability measured in the neutron star mergers is sensitive to the position of the topology change and hence offers a possibility to fix the value of $n_{1/2}$ if it can be tightened to a precise value. We address this issue here.
\begin{figure}[!h]
 \begin{center}
   \includegraphics[width=8cm]{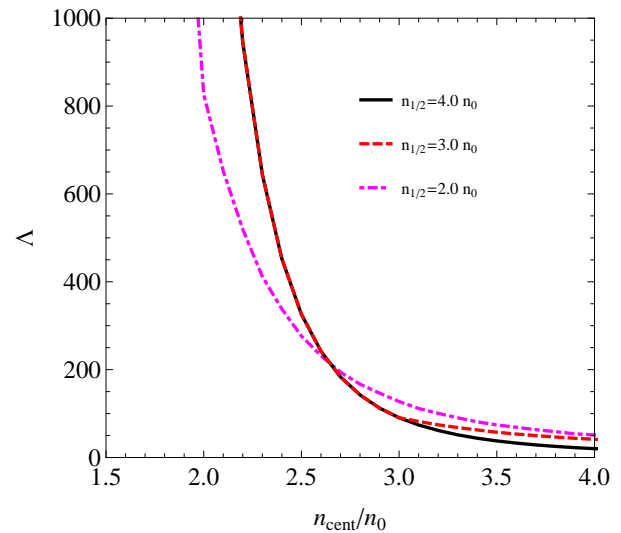}
  \end{center}
 \caption{The tidal deformability vs the central density.}
\label{fig:LvsNc}
\end{figure}

\begin{figure}[!h]
 \begin{center}
   \includegraphics[width=8cm]{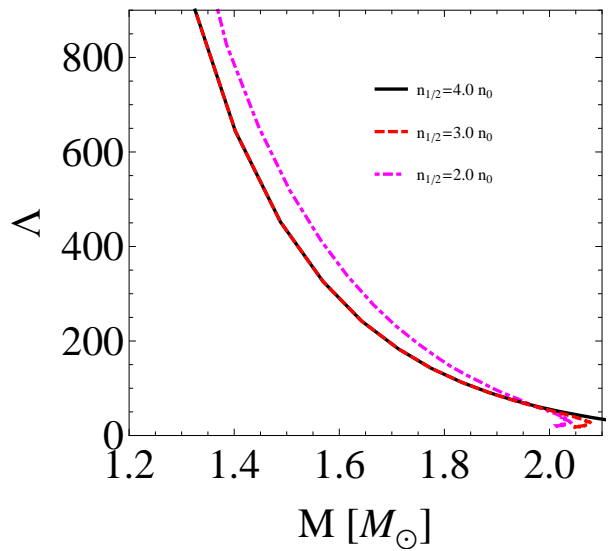}
  \end{center}
 \caption{The tidal deformability as a function of the star mass.}
\label{fig:LvsM}
\end{figure}

The  dimensionless tidal deformability parameter $\Lambda$ predicted for $n_{1/2}/n_0= 2.0, 3.0$ and $4.0$ is plotted in Figs.~\ref{fig:LvsNc} and~\ref{fig:LvsM}, the former vs the central density $n_{\rm cent}$ and the latter vs the star mass $M$. In Table \ref{tab:StarProperty} they are summarized within the range of star masses relevant to the LIGO/Virgo observation~\cite{GW170817}, together with the radii involved.

\begin{widetext}
\begin{table*}[!t]
\centering
\caption{Properties of compact stars with different masses and $n_{1/2}/n_0$.}
\label{tab:StarProperty}
\begin{tabular}{c|ccc|ccc|ccc}
\hline
\hline
\multirow{2}{*}{$M/M_\odot$}& \multicolumn{3}{c|}{$n_{cent}/n_0$}&\multicolumn{3}{|c}{ $\Lambda/100$}&\multicolumn{3}{|c}{ $R$/km}\cr\cline{2-10}
&$n_{1/2}=2.0$&$n_{1/2}=3.0$&$n_{1/2}=4.0$&$n_{1/2}=2.0$&$n_{1/2}=3.0$&$n_{1/2}=4.0$&$n_{1/2}=2.0$&$n_{1/2}=3.0$&$n_{1/2}=4.0$
\cr
\hline
1.12 &1.81&2.00&2.00&25.3&22.5&22.5&12.7&12.6&12.6\cr\hline
1.22 &1.88&2.10&2.10&16.7&14.2&14.2&12.8&12.7&12.7\cr\hline
1.31 &1.95&2.20&2.20&11.6&9.50&9.50&12.9&12.8&12.8\cr\hline
1.40 &2.02&2.30&2.30&7.85&6.52&6.52&13.0&12.8&12.8\cr\hline
1.49 &2.17&2.40&2.40&5.54&4.50&4.50&13.1&12.8&12.8\cr\hline
1.57 &2.31&2.50&2.50&4.00&3.25&3.25&13.1&12.8&12.8\cr
\hline
\hline
\end{tabular}
\end{table*}
\end{widetext}

There are striking differences between the results of $n_{1/2}=2.0n_0$ and those of $n_{1/2} > 2.0 n_0$. The former, which, we suggested, is disfavored  by the observation of the bound $\Lambda < 800$, differs appreciably from the latter, all of which share nearly the same properties of $M$ vs $n_{\rm cent}$, $\Lambda$ and the radius $R$. Note that the radius is remarkably independent of $n_{1/2}$ as well as of $M$ for $n_{1/2} > 2.0 n_0$. The only clear difference in the latter is the maximum star mass, which tends to be bigger, the greater the transition density, reaching $M_{\rm max}\sim 2.3 M_\odot$. This indicates higher $n_{1/2}$ gives higher $m_{\rm max}$ but note however that $n_{1/2}=4n_0$ is disfavored --- although not ruled out --- by heavy-ion data for the pressure $p(n)$ for $n\gsim 3n_0$.

Now turning to the case of the $\Lambda$ for the $1.4 M_\odot$ star~\cite{GW170817}  that we will denote as $\Lambda_{1.4}$, what transpires from going from $n_{1/2}=2.0n_0$ to $n_{1/2} > 2.0n_0$ is that the $\Lambda_{1.4}$ does drop significantly from $\sim 800$ to $\sim 650$. However it seems to saturate to $\sim 650$ for all $n_{1/2}$ as long as  $n_{1/2} > 2.0n_0$. This is quite reasonable --- within the framework of the present theory --- in that $\Lambda_{1.4}$ probes the density regime RI lying below the density at which the putative change of degrees of freedom takes place, that is, below where the hidden symmetries of QCD emerge. This means that $\Lambda_{1.4}$ is ignorant of, or insensitive to, the pseudoconformal structure that figures at $n\geq n_{1/2}$.

An important issue arises with this result.  A more recent analysis of \cite{GW170817} indicates~\cite{abottetal,Tsang-Press}  that the dimensionless $\Lambda$ could be tightened to a lower bound, $\Lambda = 300_{-230}^{+420}$ when mass-weighed or to $\Lambda=190_{-120}^{+390}$ with $R=11.9 \pm 1.4$~km when the same EOS is used for the stars considered. Now should the bound go down considerably lower than, say, $\sim 650$ predicted by the theory (Table \ref{tab:StarProperty}), this could not be accommodated by simply changing the topology change density. Given that the $M_{1.4}$ probes the RI regime, this would require further fine-tuning of the parameters in RI without disturbing the (good) properties of the equilibrium nuclear matter.  Whether this is feasible or not needs to be seen. But it will not affect the pseudoconformal structure that gives rise to the sound velocity $v_s^2/c^2=1/3$. A remark relevant to this issue is made below.

\section{Discussions and perspectives}
\label{last}

\label{sec:dis}

 Stated in brief, the principal result of this work is as follows. With flavor local symmetry and scale symmetry,  invisible in QCD in the matter-free vacuum, and topology intrinsic in baryonic structure implemented, we have developed the idea that the hidden symmetries together with topology change could be revealed at high density commensurate with the density of compact stars. What plays a crucial role there is the emergence of parity doubling together with what we identify as the presence of a dilaton-limit fixed point. The prediction of the approach is that there be pseudo-conformal symmetry emerging at a density at which topology change takes place in baryonic matter. Taking into consideration the recent observation of the tidal deformability in the gravitational waves from coalescing neutron stars, the pseudoconformal structure is found to set in at $n > 2n_0$ with the sound velocity converging to $v_s^2/c^2\approx 1/3$ and staying until the dilaton-limit fixed point estimated to be $> 20n_0$.

 The bound for the dimensionless tidal deformability $\Lambda$ inferred from the recent LIGO/Virgo gravitation wave suggests that the topology change cannot take place at a density less than $2n_0$ and hence sets the {\it lower bound}  $n_{1/2} > 2.0n_0$. Information from heavy-ion collisions indicates further  that the topology change cannot take place at $4n_0$, so that sets the {\it upper bound} $n_{1/2} < 4 n_0$.  This gives the bound (\ref{bound}) as announced.

 The pseudoconformal sound velocity (\ref{pc-sound}), predicted to set in at $n \gsim n_{1/2}$, typically at $n\sim 3n_0$, is totally at variance with all other predictions found in the literature. Since it is argued that such a sound velocity cannot be converged to unless there is a change of degrees of freedom from hadronic to QCD degrees of freedom~\cite{tews}, our result based on topology change can be interpreted as capturing the baryon-quark continuity which is thought to take place in the same range of density~\cite{baymetal}.

 Although much effort was devoted throughout the paper to clearly distinguish our approach from other approaches~\cite{Quarks}  where quark degrees of freedom are {\it explicitly} implemented, it is perhaps worth repeating the key point of our approach. In our approach the change of degrees of freedom, namely the topology change, does not have order parameters signaling the changeover from before to after $n_{1/2}$. In this sense the ``phase" change does not belong to the usual Landau-Ginzburg-Wilsonian paradigm. Furthermore the appearance of the precocious (pseudo)conformal sound velocity is in some sense natural in the model. We find our scenario to resemble that given by the quarkyonic matter which appears at a similar density regime~\cite{mclerran}.  Given that half-skyrmions for density $n>n_{1/2}$ are confined into baryons, the half-skyrmion matter is a baryonic matter. It seems plausible that the quarkyonic matter is also a baryonic, {\it albeit} modified, matter as has been hinted at by Ref.~\cite{philipsen}. However our picture seems to predict a (pseudo)conformal sound velocity at a density (perhaps) much lower than that of quarkyonic~\cite{vs-td}. This may be due to the nature of large $N_c$ approximations involved in the two approaches.

What is highly surprising is that the extremely simple PCM with the precocious sound velocity and a hint for emerging symmetries fully captures the full $V_{lowk}$RG physics in giving global overall accounts of compact-star properties more or less correctly. As far as we can see, there is no contradiction between the pseudoconformal structure we are advocating and the overall star observables so far available. On the other hand, most of the currently ``successful" standard nuclear models, e.g., energy-density functional approach, that carry no {\it precocious} conformal sound velocity do not seem to get into serious conflict  with the currently available observations. So the crucial question  is, is the pseudoconformal sound velocity a quantity that is indispensable for the EOS? If not, is the possible emergence of hidden symmetries,  which logically lead to the conformal sound velocity, an unphysical phenomenon? This question raises inevitably the issue as to whether the sound velocity is  a measurable quantity.

\acknowledgments

We are very grateful to Yongseok Oh for arranging our visit to the Asian Pacific Center for Theoretical Physics (APCTP) in Pohang, Korea and to the APCTP where the main part of this work was done for the hospitality. We would like to thank Hyun Kyu Lee and Won-Gi Paeng for valuable discussions and help on various important aspects of the calculation performed in this work. Y.-L. M. was supported in part by National Science Foundation of China (NSFC) under Grants No. 11475071 and No. 11747308 and the Seeds Funding of Jilin University.


\appendix

\section*{Appendix: Behavior of the Dilaton Condensate at $n\geq n_A$}

We found in Sec.~\ref{paritydoubling} that, in the mean field, the dilaton condensate drastically changes at $n_A$. Here we show that this  behavior is caused by the change of the solution for Eq.~(\ref{gapchi}) as
\begin{equation}
\bar{\chi}:\, \bar{\chi}_+ \rightarrow \bar{\chi}_-
\end{equation}
at $n \sim n_A$, where $\bar{\chi}_{\pm}$ are given by
\be
&&\frac{m_\chi^2}{f_{\chi}^2}\bchi^3_{\pm}\bar{\Sigma}_\pm
= \frac{3 m_N}{4 f_\chi} n \nonumber\\
&&\qquad\qquad\qquad\; {} \pm n\sqrt{\left(\frac{3 m_N}{4 f_\chi}\right)^2 -2 \frac{m_\chi^2}{a_\omega f_{\pi}^2}
\bar{\Sigma}_\pm \left( g_{\omega}^\ast - 1\right)^2}\,
\nonumber\\
\label{EoM3}
\ee
with $\bar{\Sigma}_\pm = \left|\ln\left(\bchi_{\pm}^2/f_\chi^2\right)\right|$ which are solutions to Eq.~(\ref{gapchi}) in the approximation that $p_F / m_N$ is small. The drastic change occurs at $n \sim n_A$, where $\bar{\chi}_+ = \bar{\chi}_-$ and the quantity $\bar{\chi}$ follows the behavior of $\bar{\chi}_-$ after $n \sim n_A$.

One can readily understand the above interplay between the nucleon mass and the $\omega$-$NN$ coupling after $n \sim n_A$. The behavior of $\bchi$ depends on how the product $(g_{v\omega}^\ast-1)^2 n^2$ goes with density. { If we expand the solution $\bar{\chi}_-$ in terms of $R(n)$ defined as
\begin{equation}
R(n) \equiv 2 \frac{m_\chi^2}{a_\omega f_{\pi}^2} \ln \left( \frac{f_\chi}{\bar{\chi}_-} \right)^2 \left( g_{v\omega}^\ast - 1 \right)^2\left( \frac{3m_N}{4f_\chi}\right)^{-2}\,,
\end{equation}
$\bar{\chi}_-$ is simplified to
\begin{equation}
\bchi_-^3 = \frac{4}{3}\frac{f_\chi^3}{a_\omega f_{\pi}^2 m_N}
\left(g_{v\omega}^\ast -1\right)^2 n + {\cal O}\left( R(n) \right)\,
\end{equation} at intermediate density, $n > n_A$, where $R(n) < 1$.}
Consequently, if $g_{v\omega}^\ast$ is constant, i.e., $B=0$, the VEV goes like $\bchi \sim n^{1/3}$. Whereas when the effective coupling varies with density as
$(g_{v\omega}^\ast-1)^2 \sim 1/n$, one finds $\bchi \sim$ const as
well captured in Fig.~\ref{mass_coupling}.

As stressed we do not expect the DLFP to be on top of chiral restoration or of the VM manifestation (in the chiral limit), but it may be close to it. So an interesting question is whether our mean-field model can say something about the chiral restoration transition.

With the conformal compensator prescription, the $\omega$-meson mass term in the present context carries $\chi^2$. This appears in the mean-field thermodynamic
potential~(\ref{omega}) in the form of $(g_{v\omega}^\ast -1)^2 n^2/\chi^2$
by use of the equation of motion for $\omega_0$. Once the density is turned
on, the inverse power of $\chi$ generates a divergent contribution to the
entire $\Omega$ peaked at $\chi=0$. This huge barrier prevents the VEV $\bchi$
from approaching the scale-symmetry (or equivalently chiral-symmetry) restored state, $\bchi= 0$. In order to have the flat $\chi$ after the onset density $n_A$ up to some density $n_B$ and then have it drop to zero for chiral restoration, some sort of level crossing must take place between the $\chi$'s as the density is increased. This may be related to the still-open problem of low-mass scalars in nuclear and hadron physics {\it vis-\`a-vis} with $f_0(500)$.

\end{document}